\documentclass[twocolumn,
               showpacs,
               preprintnumbers,
               nofootinbib,
               prd,
               superscriptaddress,
               10pt,
               notitlepage,
               aps]{revtex4-2}
\usepackage{graphicx,amssymb,amsmath,amsthm,amsfonts,epsfig,mathtools}

\usepackage[utf8]{inputenc}
\usepackage{graphicx}
\usepackage{dcolumn}
\usepackage{bm}
\usepackage{amsmath}
\usepackage{color}
\usepackage[dvipsnames]{xcolor}
\usepackage{hyperref}
\hypersetup{colorlinks=true, citecolor=MidnightBlue,linkcolor=CornflowerBlue, urlcolor=CornflowerBlue, linktocpage=true}
\usepackage{xfrac}

\newcommand{\dx}{{\rm d}^4x \sqrt{-g}}
\newcommand{\mv}{m_{\rm V}}
\newcommand{\lm}{{\ell m}}

\begin{document}

\title{Instability of vectorized stars}

\author{Ekrem S. Demirbo\u{g}a}
\email{edemirboga17@ku.edu.tr}

\author{Andrew Coates}
\email{acoates@ku.edu.tr}

\author{Fethi M. Ramazano\u{g}lu}
\email{framazanoglu@ku.edu.tr}
\affiliation{Department of Physics, Ko\c{c} University,
Rumelifeneri Yolu, 34450 Sariyer, Istanbul, Turkey}

\date{\today}

\begin{abstract}
In recent papers it has been shown that a large class of vectorization mechanisms in gravity, which involve the vector fields becoming apparently tachyonic in some regime, are actually dominated by ghosts and non-perturbative behavior. Despite this, vectorized compact object solutions have previously been found, which raises the question of how, and if, the newly discovered ghosts are quenched in these cases. Here we develop the tools to study the perturbations of vectorized compact objects, and demonstrate that they suffer from ghosts and gradient instabilities as well. Thus, these vectorized objects do not represent the stable end point of a quenched instability unlike their scalarized counterparts in the spontaneous scalarization literature.  
\end{abstract}

\maketitle

\section{Introduction}

General Relativity (GR) is incredibly successful from a phenomenological point of view, having accounted for previously unexplained observations and made many incredible predictions which have since been verified. Perhaps the most spectacular of these is the confirmation of GR's prediction of the existence of gravitational waves. It matches observations of binary pulsars so well as to provide an extremely stringent constraint on massless (and sufficiently light) scalar fields \cite{Freire:2012mg}. It is also so far consistent with all of  direct gravitational wave observations performed by the LIGO-Virgo-KAGRA collaboration (e.g \cite{LIGOScientific:2020tif}).

Given the success of GR, especially in the solar system \cite{Bertotti:2003rm}, the most interesting theories come with some mechanism to mimic GR in the solar system, while allowing for large deviations elsewhere. Such mechanisms are termed ``screening mechanisms" and the prototypical examples are the Damour Esposito-Far\`{e}se (DEF) model of spontaneous scalarization \cite{Damour:1993hw, Damour:1996ke} and the Vainshtein mechanism \cite{Vainshtein:1972sx} (see Ref. \cite{Babichev:2013usa} for a review). We are interested in the former in this study, which has an action of the form
\begin{align}
S &= \frac{1}{16 \pi} \int \dx\, \left[ R
- 2\left(\partial_\mu\phi\right)\left(\partial^\mu\phi\right)
- 2V(\phi) \right]
\nonumber \\
&\quad+ S_{\rm m} \left[ f_{\rm m};\, {\Omega}_\phi^2(\phi) \, g_{\mu\nu}\right]\ .
\label{eq:action_phi}
\end{align}
Here $V(\phi)$ is a potential for the scalar field, $f_{\rm m}$ stands for any matter fields and \(\Omega_\phi^2(\phi)\) is a conformal factor through which the scalar field and matter are coupled. Given some conditions on \(\Omega_\phi\) and \(V\) one can arrange for the theory to be solved by GR solutions with constant \(\phi =\phi_0\). Perturbations of the scalar field around such a solution have an effective mass squared of the form
\begin{equation}\label{eq:phi_mass}
   m_{\rm eff}^2= m_\phi^2- 4\pi\beta T,
\end{equation} 
where \(m_\phi^2 = \left.V''\right|_{\phi_0}\), \(\beta = \sfrac{{\rm d}^2 \log \Omega_\phi^2}{{\rm d} \phi^2}|_{\phi_0} \) and \(T\) is the trace of the stress energy tensor. Choosing \(m_\phi\) and \(\beta\) correctly one can then arrange that the solar system is stably given by GR while (say) neutron stars become unstable to scalar perturbations. An instability due to a wrong sign of the effective square mass is termed ``tachyonic". When this effective square mass is constant throughout spacetime, it is sufficient that it be negative to trigger an instability, whereas in this case it depends on the matter configuration. 

If $m_{\rm eff}^2<0$ in some region of characteristic size $R$, a heuristic condition for the instability is that $1/R^2 + m_{\rm eff}^2 \lesssim 0$ (so that some mode which can probe this length scale is unstable within it)~\cite{Ramazanoglu:2016kul}. When $m_\phi^2$ is small, one can see that the controlling parameter for the onset of the instability will be the compactness of the object. When the mass is sufficiently small, binary pulsar tests heavily constrain the model \cite{Antoniadis:2013pzd}, and for very large masses the mechanism is never activated for astrophysical objects~\cite{Ramazanoglu:2016kul}. However, there remains an interesting range of parameters for light scalars in various astrophysical scenarios~\cite{Barack:2018yly}.

There is no explicit role of the scalar nature of the field in the spontaneous scalarization mechanism we have presented, hence, there have been efforts to generalize it to other fields, the most commonly studied case being vectors~\cite{Ramazanoglu:2017xbl,Annulli:2019fzq,Ramazanoglu:2019gbz,Oliveira:2020dru}. The common theme of these theories is that the vector field spontaneously grows in the vicinity of compact objects similarly to the scalar of the DEF model, and the vectorized objects generically feature large deviations from GR.

Our analysis will focus on the direct generalization of the DEF model, Ref.~\cite{Ramazanoglu:2017xbl}, though some of our results will cover any model of vectorization that have perturbation equations of the form
\begin{equation}\label{eq:general_form}
    \nabla_\mu F^{\mu\nu} = \Xi^{\nu}{}_{\mu}A^\mu \ ,
\end{equation}
such as Refs.~\cite{Annulli:2019fzq,Ramazanoglu:2019gbz,Oliveira:2020dru}.

It was recently shown that theories of this type possess ghost instabilities if they contain what one might term a ``naive tachyon" \cite{Silva:2021jya,Garcia-Saenz:2021uyv}.\footnote{See Ref.~\cite{Esposito-Farese:2009wbc} for a similar earlier work on cosmology.} If \({\Xi}\) were to be a good mass squared tensor, it should behave, in some key ways, like $\mu^2 \delta^{\mu}{}_{\nu}$, the mass term of the Proca field. A ``naive tachyon" would then be when it does not meet these requirements, for example, if ${\Xi}$ has negative eigenvalues as a matrix, one would expect some tachyonic behaviour. As we will demonstrate later, ${\Xi^{\mu\nu}}$ plays an even more prominent role than that of a mass squared tensor, being the inverse of the effective metric for the scalar mode of the vector field, and so requirements of causality will also play a role in deciding what forms of  ${\Xi}$ allows healthy dynamics of the vector (see Ref.~\cite{Babichev:2018uiw}).

As a concrete example, the spontaneous vectorization theory of Ref.~\cite{Ramazanoglu:2017xbl} is given by the action\footnote{Also see Ref.~\cite{BeltranJimenez:2013fca} for a similar cosmological theory.}
\begin{align}
S &= \frac{1}{16 \pi} \int \dx\, \left( R
- F_{\mu\nu} F^{\mu\nu}
- 2 \mv^2 A_{\mu}A^{\mu} \right)
\nonumber \\
&\quad+ S_{\rm m} \left[ f_{\rm m};\, {\Omega_{\rm V}}^2(A_{\mu}A^\mu) \, g_{\mu\nu}\right],
\label{eq:action}
\end{align}
where $A_{\mu}$ is a vector field with bare mass $\mv$, $F_{\mu\nu} = \nabla_\mu A_\nu - \nabla_\nu A_\mu$, and as before $f_{\rm m}$ is a generic label for any matter field. The vector field equation of motion is given by
\begin{equation}
    \nabla_{\mu} F^{\mu\nu} = \left(\mv^2 - 8 \pi \Lambda \, \Omega_{\rm V}^{4} \tilde{T} \right) A^{\nu}
    \equiv \hat{z} \mv^2 A^\nu\ ,
    \label{eq:eom_vt}
\end{equation}
where
\begin{equation}
    \hat{z} = 1- \frac{8\pi}{\mv^2} \frac{d \ln \Omega_{\rm V}}{d(A_{\mu}A^{\mu})} \Omega_{\rm V}^4 \tilde{T}   \ .
    \label{eq:zhat_def}
\end{equation}
$\tilde{T}$ is the trace of the stress energy tensor with respect to the metric $\tilde{g}_{\mu\nu} = \Omega_{\rm V}^2 g_{\mu\nu}$. In the rest of the paper we will use 
\begin{equation}
    \Omega_{\rm V} = e^{\beta A_\mu A^\mu /2}\ ,
\end{equation}
following the literature.

When $\hat{z}<0$ we have a naive tachyon as described above, which led to expectations that the theory would have vectorized neutron star solutions similar to the scalarized solutions of the DEF model~\cite{Ramazanoglu:2017xbl}. However, in Ref.~\cite{Silva:2021jya} it was shown that  action~\eqref{eq:action} also features ghost instabilities for a neutron star that is a solution of GR when $\hat{z}$ is negative. Furthermore, $\hat{z}$ also changes sign in many cases, which leads to other serious problems due to divergent terms of the form $\hat{z}^{-1}$.

Despite the problems around a GR background, neutron star solutions with non-trivial vector configurations have been numerically calculated~\cite{Ramazanoglu:2017xbl}. If these vectorized solutions are stable, one could hope that the problematic GR backgrounds are dynamically avoided and these vectorized configurations develop without issue. Here, we will show that this is not the case. That is, arbitrarily small perturbations from vectorized solutions calculated in Ref.~\cite{Ramazanoglu:2017xbl} grow in an unbounded manner, meaning they are unstable. We also explore parts of the $(\beta,\mv)$ parameter space that have not been investigated so far, and show that there is no stable spherically symmetric vectorized star in any sector we studied. This is in stark contrast to the case of the DEF model where typical scalarized solutions are known to be stable in various models~\cite{Harada:1998ge,Salgado:1998sg,Novak:1998rk,Mendes:2016fby,Ramazanoglu:2016kul,East:2021bqk}.

Despite these negative results, we shall see that the naive picture of a tachyonic degree of freedom can apply to the axial part of the vector field, and so, in the regions of parameter space where this holds, one should likely be looking at axisymmetric stars for healthy end states, if any exist.

In Sec.~\ref{sec:instabilities}, we develop the perturbative framework to analyze the instabilities of vectorized stars by generalizing the approach of Ref.~\cite{Silva:2021jya}. In Sec.~\ref{sec:results}, we numerically solve for vectorized stars in the theory of action~\eqref{eq:action}, and show that all solutions are unstable to linearized pertubations. Not all solutions carry all the different forms of instabilities we study, but they all carry at least one. In Sec.~\ref{sec:conclusions}, we will discuss our results, and argue how our methods can be applied to other vectorization theories and other extensions of the DEF model. 

We use the geometric units $G=c=1$ and the metric signature $(-1,1,1,1)$. We use Einstein summation conventions for greek indices $\mu,\nu, \dots$, but not for the latin indices for spatial variables such as $i$, unless specifically stated otherwise.

\section{Instabilities of vectorized stars}
\label{sec:instabilities}
\subsection{Spherically symmetric vectorized stars}
The metric of a static and spherically symmetric vectorized star can be represented by the ansatz
\begin{equation}
{\rm d} s^2 = - e^{\nu(r)} {\rm d} t^2 + \frac{1}{1-\frac{2\mu(r)}{r}} {\rm d} r^2 +
r^2 ({\rm d} \theta^2 + \sin^2\theta\,{\rm d}\phi )\ .
\label{eq:line_element}
\end{equation}
The vector field also has the simple form
\begin{align}
    A^\mu = \left( A^0(r),0,0,0 \right) = A^0(r) \partial_0\ .
    \label{eq:vec_solution}
\end{align}
That is, only the time component of the vector field is nonvanishing in a static and spherically symmetric spacetime. This was ``assumed'' in Ref.~\cite{Ramazanoglu:2017xbl} for ease of calculation. Here, we examine the most general case, and show that there is no need for an assumption.

The fact that spherical symmetry implies that the angular components of the vector field, $A^{\theta,\phi}$, vanish is not a trivial one. For example, minimally coupled massless vector fields can have nonzero angular components under spherical symmetry such as the electromagnetic field vector $V$ of a Reissner-Nordstrom black hole with magnetic charge. This is possible due to the fact that the stress-energy tensor can still be spherically symmetric in this case, even though $V^\phi \neq 0$ and $F^{\theta\phi} \neq 0$. However, this is not the case for a massive field.

The metric equation of motion arising from the action~\eqref{eq:action} takes the following form outside the star
\begin{align}
    R_{\mu\nu} =
    2 F_{\mu\rho} F_{\nu}{}^\rho -\frac{1}{2} F_{\rho\sigma} F^{\rho\sigma} g_{\mu\nu}
    +2\mv^2 A_\mu A_\nu \ .
    \label{eq:einstein_eq}
\end{align}
The line element~\eqref{eq:line_element} has direct implications for the Ricci and Einstein tensors. Firstly, $R_{\mu\nu}$ has to be diagonal, hence only one component of $A_\mu$ can be nonvanishing due to the $\mv^2$ term. Secondly, the surviving component cannot be $A^\theta$ or $A^\phi$, since Eq.~\eqref{eq:line_element} also implies that $R_{\phi\phi} = \sin^2 \theta R_{\theta\theta}$, which cannot be satisfied due to the same term. 

Finally, vanishing of $A^r$ is a direct consequence of the $\alpha=r$ case of
Eq.~\eqref{eq:eom_vt}
\begin{align}
A^r= \frac{\partial_\rho (\sqrt{|g|} F^{\rho r}) }{\sqrt{|g|} (-4\pi \Omega_{\rm V}^4 \beta \tilde{T} +m^2 )}
=0 \ ,
\end{align}
where we used the partial derivative formula for the divergence of an antisymmetric tensor, and the fact that the only nonzero component of the differentiated tensor, $F^{tr}$, is time independent for a static solution. Overall, the only surviving component is $A^0$.

In the following, we will assume that the nuclear matter behaves as a perfect fluid with stress-energy tensor
\begin{equation}\label{fluid_eq}
\tilde{T}^{\mu\nu}=(\tilde{\rho}+\tilde{p})\tilde{u}^{\mu}\tilde{u}^{\nu}+\tilde{p} \tilde{g}^{\mu\nu}\ ,
\end{equation}
where the total energy density $\tilde{\rho}$, pressure $\tilde{p}$ and the components of the fluid 4-velocity $\tilde{u}^{\alpha}$ only depend on the radial coordinate $r$.

\subsection{Basics of tachyons, ghosts and gradient instabilities}
Although they are related to each other, we will categorize our instabilities into three groups: tachyons, ghosts and gradient instabilities. Their basic behavior can be demonstrated for a classical massive scalar field theory in \(1+1 D\)
\begin{align}
\label{eq:instability_action}
S_\varphi = \int {\rm d}t {\rm d}x\ \sqrt{-g}
\left[ -g^{tt} (\partial_t \varphi)^2
- g^{xx} (\partial_x \varphi)^2 - m^2 \varphi^2 \right]
\end{align}
whose equation of motion is
\begin{align}
\label{eq:instability_eom}
g^{tt} \partial_t^2 \varphi
+g^{xx}\partial_x^2 \varphi
=m^2 \varphi \
\end{align}
for a metric $g_{\mu\nu}$ with constant diagonal components.

A Fourier mode $\varphi(t,x)=e^{i\left[\omega(k)t-kx\right]}$ has the dispersion relation
\begin{align}
\omega(k) = \sqrt{ \frac{m^2+g^{xx} k^2}{-g^{tt}} }\ .
\end{align}
If $\omega(k)$ is real, the mode has an oscillatory behavior in time. This happens in the ``usual'' case where $g^{tt}<0$, $g^{xx}>0$ and $m^2>0$. If $\omega(k)$ becomes imaginary, the mode grows exponentially, which is an instability. Let us consider the cases where one of the constants change its sign while the other two have the usual one.

We call the $m^2<0$ case a ``tachyon.'' The modes with
\begin{align}
    k^2 < \frac{-m^2}{g^{xx}} 
\end{align}
are unstable in this case, hence it is an infrared instability. Note that high wave number modes are stable, and the fastest growing unstable mode behaves as $\sim e^{\sqrt{m^2/g^{tt}}\, t}$, hence, there is an upper limit to the growth rate. This is the instability that underlies the spontaneous scalarization mechanism of the DEF model~\cite{Damour:1993hw}.

We  call the $g^{tt}>0$ case a ``ghost.'' All ghost modes exponentially grow, the growth rate of the modes diverging at high wave numbers as $\sim e^{\sqrt{g^{xx}/g^{tt}}k\, t}$ . Thus, unlike a tachyon, a ghost, as described here, has infinite growth rate for generic perturbations.  

We  call the $g^{xx}<0$ case a ``gradient instability.'' The modes with
\begin{align}
    k^2 > \frac{m^2}{-g^{xx}} 
\end{align}
are unstable, and low wave number modes are stable. However, high wave number modes can grow arbitrarily fast like the ghost as $\sim e^{\sqrt{g^{xx}/g^{tt}}k\, t}$. Hence, the growth rate of this instability also diverges.

In our simple model, a coexistence of two of the instabilities is equivalent to the third instability. For example, a ghost and a gradient with $m^2>0$ behaves exactly as a tachyon, which is the simple mathematical result of the fact that we can multiply the action~\eqref{eq:instability_action} or the equation of motion~\eqref{eq:instability_eom} with $-1$, and have the same theory. Similarly, $g^{tt}>0$, $g^{xx}<0$ and $m^2<0$ can be considered as a completely stable theory. However, the scalar field couples to other fields, and the overall sign of the action is meaningful in most situations, e.g. in spontaneous scalarization in Eq.~\eqref{eq:action_phi}. Hence, we will categorize each sign change as a separate instability.

Lastly, we will see that the instabilities we encounter will have $g_{\mu\nu}$ and $m^2$ values that are functions of spacetime coordinates, and they can have different signs in different regions. In the case of the tachyon, having $m^2<0$ only in a finite region of space effectively puts a lower bound on the $k$ values that can be tachyonic, which might mean no mode is tachyonic~\cite{Ramazanoglu:2016kul}. On the other hand, having different regions of spacetime with positive and negative $g^{tt}$ values means the coefficient of the principal part of the equation of motion~\eqref{eq:instability_eom} vanishes at the transition points between regions. Alternatively, if we divide the whole equation by $g^{tt}$, some of the coefficients diverge at such points. In any case, perturbative treatment of such an equation becomes impossible, which we will see to be the case for vectorization.

\subsection{Ghost and gradient instabilities in vectorization}
The stability analysis for any putative solution to the vectorization theory can start at the generalized Lorenz condition~\cite{Silva:2021jya} 
\begin{align}
    \nabla_\mu (\hat{z} A^\mu) = 0
    \label{eq:lorenz}
\end{align}
which is obtained from Eq.~\eqref{eq:eom_vt} by realizing that $\nabla_\mu\nabla_\nu F^{\mu\nu} = 0$ due to the antisymmetry of $F^{\mu\nu}$. Let us fix the metric of the vectorized solution as the background metric, and consider perturbative deviations of the vector field
\begin{align}
    A^\mu = \bar{A}^\mu + \delta A^\mu\ .
\end{align}
Here, and from now on, an overbar denotes the value of a function or operator for the exact vectorized solution.
The linearized Lorenz condition becomes
\begin{align}
    \overline{\nabla}_\mu \left(\Xi^{\mu}{}_\nu \delta A^\nu \right) = 0\ ,
    \label{eq:lorenz_linear}
\end{align}
where we define
\begin{align}
    \Xi^\mu{}_\nu \equiv \bar{z}\ \delta^\mu{}_\nu + 2\  \overline{\frac{d\hat{z}}{d(A^\rho A_\rho)}}  \ \bar{A}^\mu \bar{A}_\nu \ .
    \label{eq:xi}
\end{align}
For the exponential conformal factor $\Omega_{\rm V} = e^{\beta A^\mu A_\mu /2}$ of Ref.~\cite{Ramazanoglu:2017xbl}
\begin{align}
    \mv^2\ \Xi^\mu{}_\nu &= -16\pi \beta^2 \tilde{T}_{\rm bg} e^{2 \beta \bar{A}^\rho \bar{A}_\rho}\ \bar{A}^\mu \bar{A}_\nu \nonumber \\
    &+ \left( \mv^2 - 4\pi \beta \tilde{T}_{\rm bg} e^{2 \beta \bar{A}^\rho \bar{A}_\rho}   \right)\ \delta^\mu{}_\nu \ ,
    \label{eq:xi2}
\end{align}
where $\tilde{T}_{\rm bg}$ is the value of the trace of the stress-energy tensor for the vectorized solution. More explicitly
\begin{align}
    \Xi^0{}_0 &= 1 - \frac{4\pi \beta \tilde{T}_{\rm bg} e^{2 \beta \bar{A}^0 \bar{A}_0} \left(1+4\beta \bar{A}^0 \bar{A}_0\right)}{\mv^2} \label{eq:xi00} \\
    \Xi^i{}_i &=  1 - \frac{4\pi \beta \tilde{T}_{\rm bg} e^{2 \beta \bar{A}^0 \bar{A}_0}}{\mv^2}\ \ \ \textrm{(no sum)}\ , 
    \label{eq:xiii}
\end{align}
where there is no summation over the repeated spatial index $i$ as is the case for most expressions in this paper. These reduce to the corresponding expressions in Ref.~\cite{Silva:2021jya} when $\bar{A}^\mu=0$, as expected.

We can already see some of the problematic aspects of the vectorized solutions directly using the Lorenz condition. The form of the metric and the vector field for the vectorized star imply that $\Xi^\mu{}_\nu$ is diagonal on this background, and Eq.~\eqref{eq:lorenz_linear} gives the time evolution for $\delta A^0$ 
\begin{align}
    \partial_0 \left( \sqrt{-\bar{g}}\ \Xi^{0}{}_{0} \delta A^0\right)
    = -\sum_{i=1}^3 \partial_i \left( \sqrt{-\bar{g}}\ \Xi^{i}{}_{i} \delta A^i \right)\ , 
    \label{eq:lorenz_partial}
\end{align}
where the summation  over $i$ is shown explicitly.\footnote{$A^0$ is not a dynamical degree of freedom since its time derivative does not appear in the action~\eqref{eq:action}, but the dynamics of the other components imply this time evolution indirectly~\cite{Heisenberg:2014rta}.} This is an advection equation for the vector density $\delta B^\mu \equiv \sqrt{-\bar{g}}\ \Xi^{\mu}{}_{\nu} \delta A^\nu$. Consider a point $p$ where $\Xi^{0}{}_{0}=0$. Note that the right hand side of Eq.~\eqref{eq:lorenz_partial} is generically nonzero at $p$ since there is no symmetry principle to ensure otherwise. This means even if the initial data is such that $\delta B(t=0)=0$ at $p$ , it will immediately evolve to nonzero values, assuming a meaningful time evolution is possible. However, this implies that the time derivative of $\delta A^0$ diverges at $p$. In summary, $\Xi^{0}{}_{0}=0$ implies divergences in the time evolution of generic pertubations around a vectorized star.

Eq.~\eqref{eq:lorenz_linear} is first order in time, hence the nature of the divergence is not apparent. We can show that this is related to a ghost instability by following Ref.\cite{Silva:2021jya} and using the Stueckelberg trick~\cite{Ruegg:2003ps}, which makes the following substitution in the action~\eqref{eq:action}
\begin{equation}
    A_\mu \to A_\mu+\frac{1}{\mv}\partial_\mu\psi \ .
\end{equation}
This separates out an explicit scalar field, whose equation of motion can be obtained by varying the action, or equivalently by using the Stueckelberg trick directly in the Lorenz condition (Eq.~\eqref{eq:lorenz}),
\begin{equation}
    \nabla_\mu \left[ \hat{z} \left( A_\mu+\frac{1}{\mv}\nabla_\mu\psi \right)\right] = 0 \ .
\end{equation}
For perturbative deviations from a vectorized star solution, the linearized equation is
\begin{equation}
    \overline{\nabla}_\mu \left[ \Xi^{\mu\nu} \left( \delta A_\nu+\frac{1}{\mv}\overline{\nabla}_\nu \delta \psi \right) \right] \ ,
\end{equation}
where
\begin{align}
    \Xi^{\mu\nu} = \overline{g}^{\nu\rho}\Xi^{\mu}{}_\rho
    = \bar{z}\ \overline{g}^{\mu\nu}
    +2\ \overline{\frac{d\hat{z}}{d(A^\rho A_\rho)}} \ \bar{A}^\mu \bar{A}^\nu \ .
    \label{eq:xi3}
\end{align}
The significance of $\Xi^{\mu\nu}$ becomes clearer when we consider the principal part, the highest derivative term, of this equation: ${\Xi}^{\mu\nu} \partial_\mu\partial_\nu \delta\psi$. This means that the evolution of $\delta \psi$ is governed by the effective inverse metric
\begin{equation}
    \breve{g}^{\mu\nu} = \Xi^{\mu\nu}\ .
\end{equation}
In particular, if \(\Xi^{00}\) is positive \(\delta\psi\) will be a ghost, and if \(\Xi^{ii}\) is negative for any $i$, \(\delta\psi\) will have a gradient instability for large wave numbers. Using this effective metric the full equation of motion can be expressed as
\begin{equation}\label{eq:psi_eom}
    \breve\Box \delta \psi = -\mv\breve{\nabla}\cdot \delta A + \Delta\Gamma^{\beta}_{\alpha\beta}\Xi^{\mu\alpha}\left(\mv \delta A_\mu+ \breve{\nabla}_\mu \delta \psi\right)
\end{equation}
where $\Delta\Gamma^{\mu}_{\alpha\beta}=\breve{\Gamma}^{\mu}_{\alpha\beta}-\Gamma^{\mu}_{\alpha\beta}$, and both $\breve\Box$ and $\breve\nabla\cdot$ imply contraction with $\breve{g}$, rather than $g$. 
Note that, $\breve{g}^{00}$ or equivalently $\Xi^{00}$ changes sign exactly when $\Xi^{0}{}_0$ changes sign, since this tensor and the metric are both diagonal in our cases of interest. Hence, the divergence in the Lorenz condition Eq.~\eqref{eq:lorenz_partial} occurs exactly when $\delta \psi$ becomes a ghost. 

Even though we considered a specific model, Ref.~\cite{Ramazanoglu:2017xbl}, so far, note that our results are valid for any symmetric $\Xi$, hence applies to other vectorization models studied in detail in Refs.~\cite{Annulli:2019fzq} and~\cite{Ramazanoglu:2019gbz}. It is, however, worth noting that non-diagonal terms, which one would expect to find outside of spherical symmetry, complicate the analysis somewhat, again see Ref.~\cite{Babichev:2018uiw}. 

Lastly, we can also observe the ghost directly in the vector field components when we recast the field equation~\eqref{eq:eom_vt} in the form
\begin{equation}
    \Box A_\alpha + \left(\nabla_\beta \ln\hat{z}\right) \nabla_\alpha A^\beta = \mathcal{M}_{\alpha\beta}A^\beta\ ,
    \label{eq:vec_hyperbolic}
\end{equation}
where we used the Lorenz constraint and the commutation rules for covariant derivatives~\cite{Silva:2021jya}. The mass squared tensor is defined as
\begin{equation}
    \mathcal{M}_{\alpha\beta}=\hat{z} m^2 g_{\alpha\beta}+R_{\alpha\beta}-\nabla_{\alpha}\nabla_{\beta}\ln\hat{z}\ .
\end{equation}

The principle part of Eq.~\eqref{eq:vec_hyperbolic} solely consists of $\Box A_\alpha$ when it is linearized around the GR solution for which the vector field vanishes. On the other hand, for a spherically symmetric vectorized star background, the following term also contributes to the linearized equation
\begin{equation}\label{eq:extraprinciplepart}
    \overline{\nabla}_{\alpha}\overline{\nabla}_{\beta}\ln\hat{z} = \frac{2}{\bar{z}}\ \overline{\frac{d\hat{z}}{d(A^\rho A_\rho)}} \bar{A}^\sigma \overline{\nabla}_{\alpha}\overline{\nabla}_{\beta} \delta A_\sigma + \dots
\end{equation}
This means the principal part can not be expressed as a wave operator in general, however, for $\alpha=0$ in Eq.\eqref{eq:vec_hyperbolic} it can be written as
\begin{equation}\label{eq:vec_ghost}
    \breve{g}^{\mu\nu} \partial_\mu \partial_\nu \delta A_0\ ,
\end{equation}
where we used the fact that only the $0$ component of $\bar{A}_\rho$ is nonvanishing. Hence, $\breve{g}^{\mu\nu}=\Xi^{\mu\nu}$ is also the effective metric that governs the evolution of $\delta A_0$, which becomes a ghost if $\breve{g}^{00}$ changes sign, and has a gradient instability if $\breve{g}^{ii}$ changes sign for any $i$. These are the same conditions for the existence of instabilities in the scalar field $\psi$ arising from the Stueckelberg trick.

To summarize, we have shown that the existence of ghost or gradient instabilities depends on the behavior of $\Xi^{\mu\nu}$, which acts as an effective (inverse) metric for some degrees of freedom in the theory of action~\eqref{eq:action}. Since $g_{\mu\nu}$ always has the signature $[-1,1,1,1]$, a change of sign in $\Xi^\mu{}_\nu$ occurs exactly under the same conditions of a change of sign of $\Xi^{\mu\nu}$ as we mentioned. Thus, we will compute  $\Xi^\mu{}_\nu$ profiles of vectorized neutron stars in Section \ref{sec:results} in order to assess their stability.

\subsection{Tachyons in vectorization}
The original aim of vectorization as introduced in Eq.~\eqref{eq:action} was to replicate the tachyon-based spontaneous growth of scalar fields arising from action~\eqref{eq:action_phi}~\cite{Ramazanoglu:2017xbl}. In the standard scenario, the stability of the final solution is realized by nonlinear effects which suppress the instability as the scalar field grows~\cite{Ramazanoglu:2016kul}.

The existence of a ghost instability for perturbations around a vectorized star is sufficient to demonstrate that such a solution is unstable, however, it is interesting to see if the suppression mechanism of the tachyonic instability also fails. Namely, it is possible that the nonlinear effects that are expected to quench the tachyon once a vectorized solution forms fail.

GR backgrounds are known to be susceptible to tachyon instabilities in vectorization~\cite{Silva:2021jya}. The existence of ghosts breaks the perturbative approach due to divergent terms in the time evolution, but let us try to see the behavior of the tachyonic modes assuming a background solution exists and we only perturb the degrees of freedom which are tachyonic. We will repeat the discussion of Ref.~\cite{Silva:2021jya} for a vectorized background, starting with a vector spherical harmonic decomposition of the perturbative deviations from the exact solution
\begin{equation}
\delta A_{\alpha} = \frac{1}{r}\sum^{4}_{i=1}\sum_\lm
c_{i} u^\lm_{(i)}(t,r) Z^{(i)\lm}_{\alpha}(\theta, \phi),
\end{equation}
where $c_1 = c_2 = 1$, $c_3 = c_4 = 1/\sqrt{\ell(\ell + 1)}$, and 
\begin{align}
Z^{(1)\lm}_{\alpha} &= [1,0,0,0]Y_\lm,
\\
Z^{(2)\lm}_{\alpha} &= [0, 1,0,0]Y_\lm,
\\
Z^{(3)\lm}_{\alpha} &= \frac{r}{\sqrt{\ell(\ell+1)}}
[0,0,\partial_{\theta},\partial_{\phi}]Y_\lm,
\\
Z^{(4)\lm}_{\alpha} &= \frac{r}{\sqrt{\ell(\ell+1)}}
[0,0,\csc\theta \partial_{\phi}, -\sin\theta\partial_\theta]Y_\lm\ ,
\end{align}
$Y_\lm$ being the scalar spherical harmonics. Details of this decomposition can be found in Refs.~\cite{Silva:2021jya} and~\cite{Rosa:2011my}.

As the background vector field $\bar{A}$ has no \(\theta\) or \(\phi\) component, the $u_4$ degree of freedom, the so-called axial mode, behaves as in the case with vanishing vector field, with a slightly different expression for the effective mass. In particular, the perturbation Lagrangian up to quadratic terms for any $u_4 \equiv u_{(4)}^{\ell 0}$ becomes 
\begin{align}
S_{\rm t} &= \int \mathrm{d} t \mathrm{d} r 
\frac{ e^{\bar{\nu}/2}\left(1-2\bar{\mu}/r\right)^{-1/2}}{4 \pi \ell(\ell + 1) }
\left\{
\vphantom{\frac{1}{2}} 
e^{-\bar{\nu}} (\dot{u}_{4})^{2}
\right.
\nonumber \\
&\quad \left.
- \left( 1 - \frac{2\bar{\mu}}{r} \right) (u'_{4})^{2}
- \left( \frac{\ell(\ell + 1)}{r^2} + \bar{z}\mv^2 \right) u_{4}^2
\right\}
\nonumber \\
\label{eq:u4_lag}
\end{align}
after integration over the angular coordinates. This leads to the field equation
\begin{align}
    \bigg[ 
    &-e^{-\bar{\nu}} \partial_t^2 + e^{-\bar{\nu}/2}\sqrt{ 1 - \frac{2\bar{\mu}}{r}} \partial_r \left(e^{\bar{\nu}/2}\sqrt{ 1 - \frac{2\bar{\mu}}{r}} \partial_r \right) \nonumber\\
    &- \frac{\ell(\ell + 1)}{r^2} - \bar{z}\mv^2 \bigg] u_4 = 0 \ .
\end{align}
If we perform a separation of variables $u_4(t,r)=e^{i\omega t} u(r)$, we obtain %
\begin{align}\label{eq:eigenfunction}
    \left[
    -\frac{d^2}{dr_*^2}
    + V_{\rm eff}(r) \right] u
    = \omega^2\ u \ ,
\end{align}
where
\begin{align}
    \frac{d r_*}{dr} &= e^{-\bar{\nu}/2}\left( 1 - \frac{2\bar{\mu}}{r} \right)^{-1/2} \\
    V_{\rm eff}(r) &= e^{\bar{\nu}} \left( \frac{\ell(\ell + 1)}{r^2} +\bar{z}\mv^2 \right)\ ,
    \label{eq:Veff}
\end{align}
and $V_{\rm eff}$ can be considered a function of $r_*$ through $r(r_*)$.
A tachyon exists if there is a solution to this eigenfunction problem that satisfies $\omega^2<0$. Note that the axial mode is only defined for $\ell \geq 1$.

Increasing $\ell$ makes the modes strictly less tachyonic due to its positive contribution to $V_{\rm eff}$, which means it is sufficient to check $\ell=1$ for the existence of a tachyon. Non-existence of linearized tachyons in the axial sector is guaranteed if $V_{\rm eff}>0$ everywhere, since the expectation value of the differential operator on the left hand side of Eq.~\eqref{eq:eigenfunction} for any trial function is positive in this case, which ensures $\omega^2>0$ for any eigenfunction. We will employ this shortcut when analyzing our vectorized stars.   

The above analysis is only for a single vector spherical harmonic, and does not have any direct implications for the even modes $u_{1,2,3}$. This means, we can conclude the existence of tachyons from Eq.~\eqref{eq:eigenfunction}, but not rule them out completely even if $u_4$ is not tachyonic. We will not investigate the more complicated coupled equations for $u_{1,2,3}$ in this study, as we will soon see that the even sector is already plagued by ghost and gradient instabilities.

\section{Results} \label{sec:results}
\begin{figure}
    \includegraphics[width=0.49\textwidth]{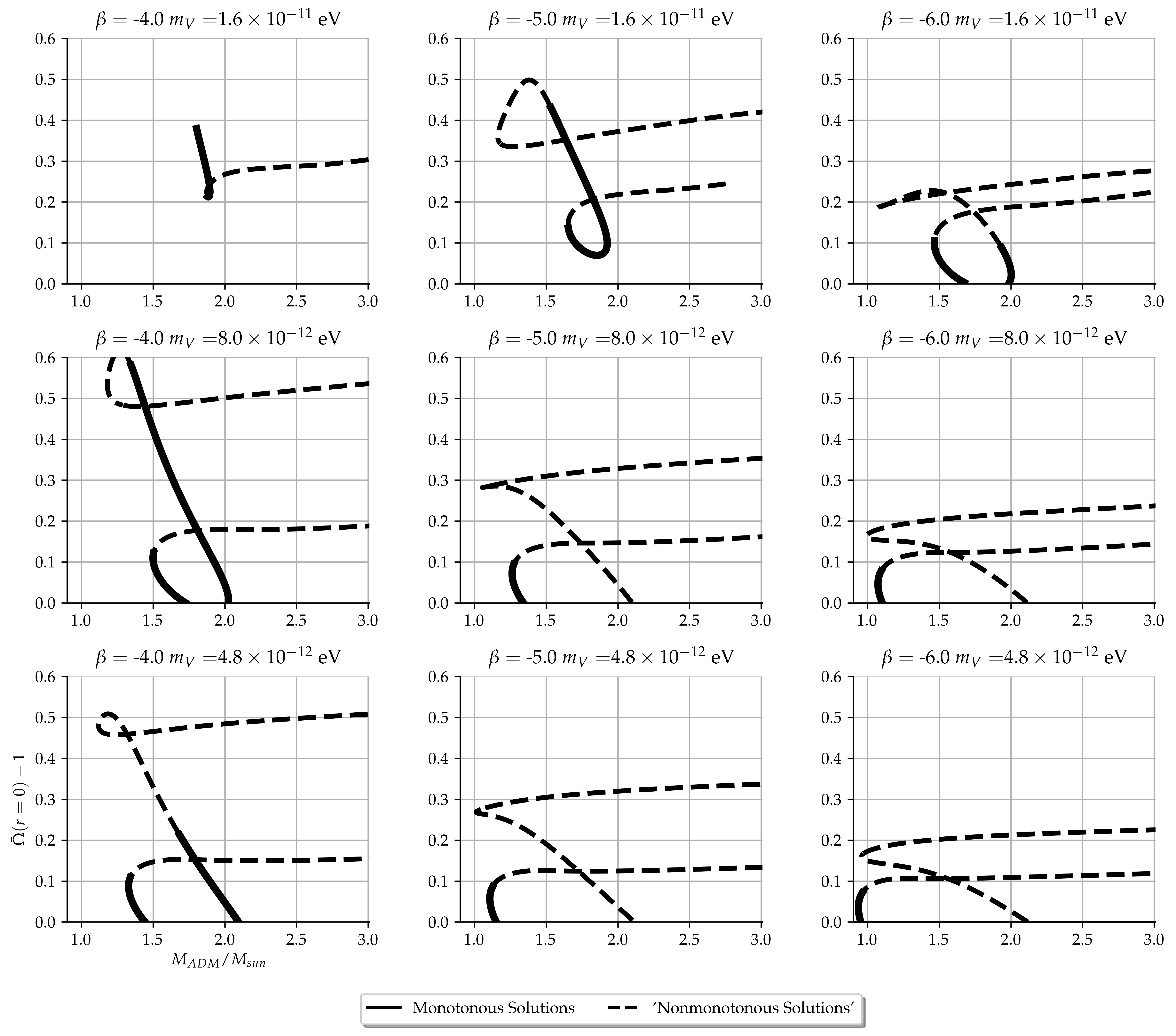}
    \caption{Deviations from GR measured by $\overline{\Omega}_{\rm V}(r=0)-1$ as a function of the ADM mass of vectorized stars for various values of $\beta<0$ and $\mv$. This is a slightly updated version of Fig.~1 of Ref.~\cite{Ramazanoglu:2017xbl}. There are solutions where the vector field or the matter density does not monotonically decrease with radius (dashed lines), and the general dependence of deviations from GR on the neutron star mass is qualitatively different from the case of scalarization. e.g. in Ref.~\cite{Ramazanoglu:2016kul}. }
    \label{fig:vector_strength_negative_beta}
\end{figure}
\begin{figure}
    \includegraphics[width=0.49\textwidth]{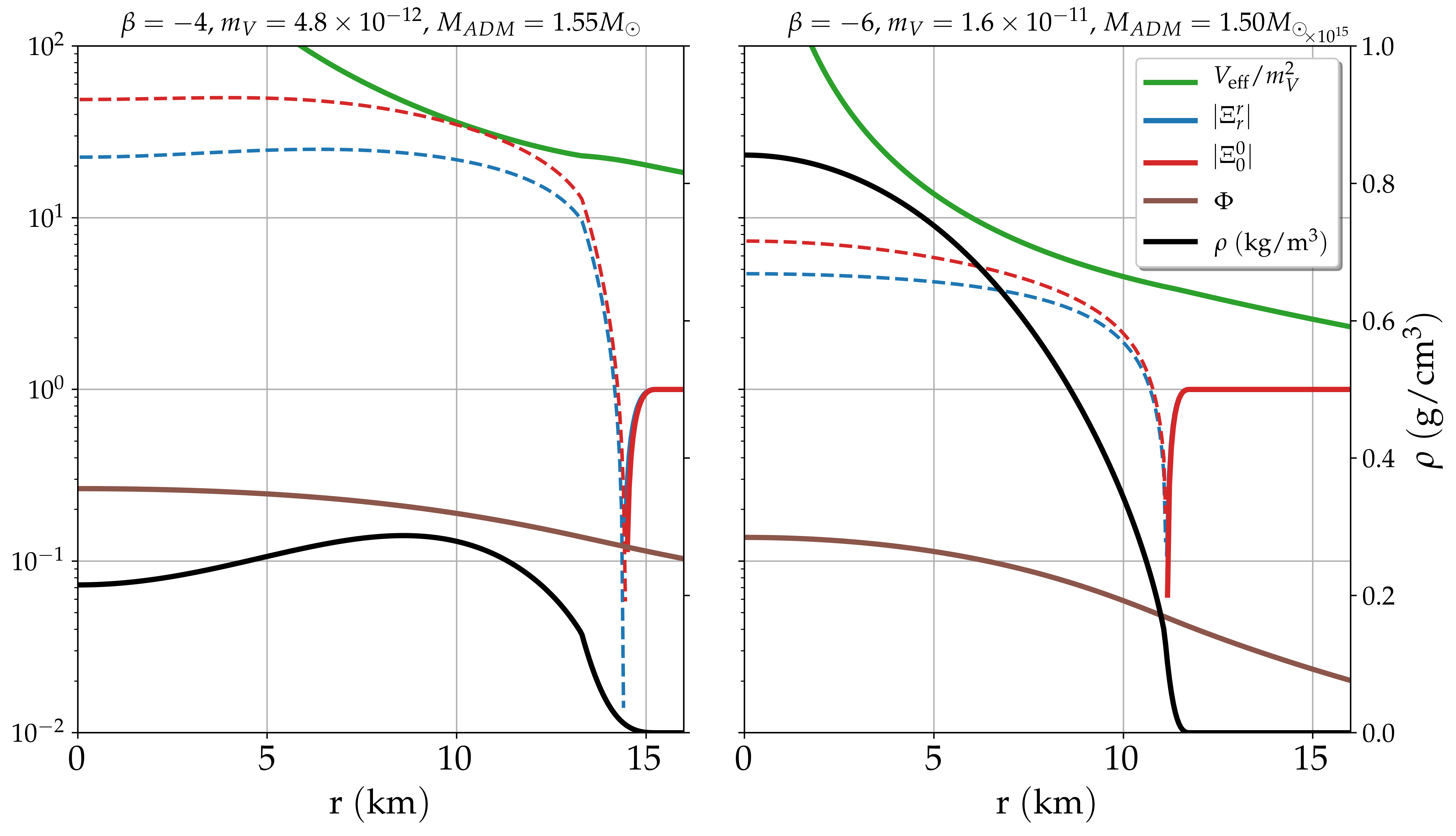}

    \caption{$\Xi^0{}_0$, $\Xi^r{}_r$, $V_{\rm eff}$, $\Phi = \sqrt{-A_0A^0}$ and the matter energy density $\tilde{\rho}$ of a non-monotonic (left) and monotonic (right) star for $\beta<0$. The radii where $\Xi^0{}_0$ and $\Xi^r{}_r$ cut through $0$ can be seen as the divergences in the logarithmic scale, and the lines are dashed when the functions attain negative values. $V_{\rm eff}$ is monotonically decreasing with radius in both cases and is positive everywhere. Note that the regions where $\Xi^0{}_0$ and $\Xi^r{}_r$ become negative overlap closely, but not exactly.
    }
    \label{fig:star_profiles}
\end{figure}
\begin{figure*}
    \centering
    \includegraphics[width=0.9\textwidth]{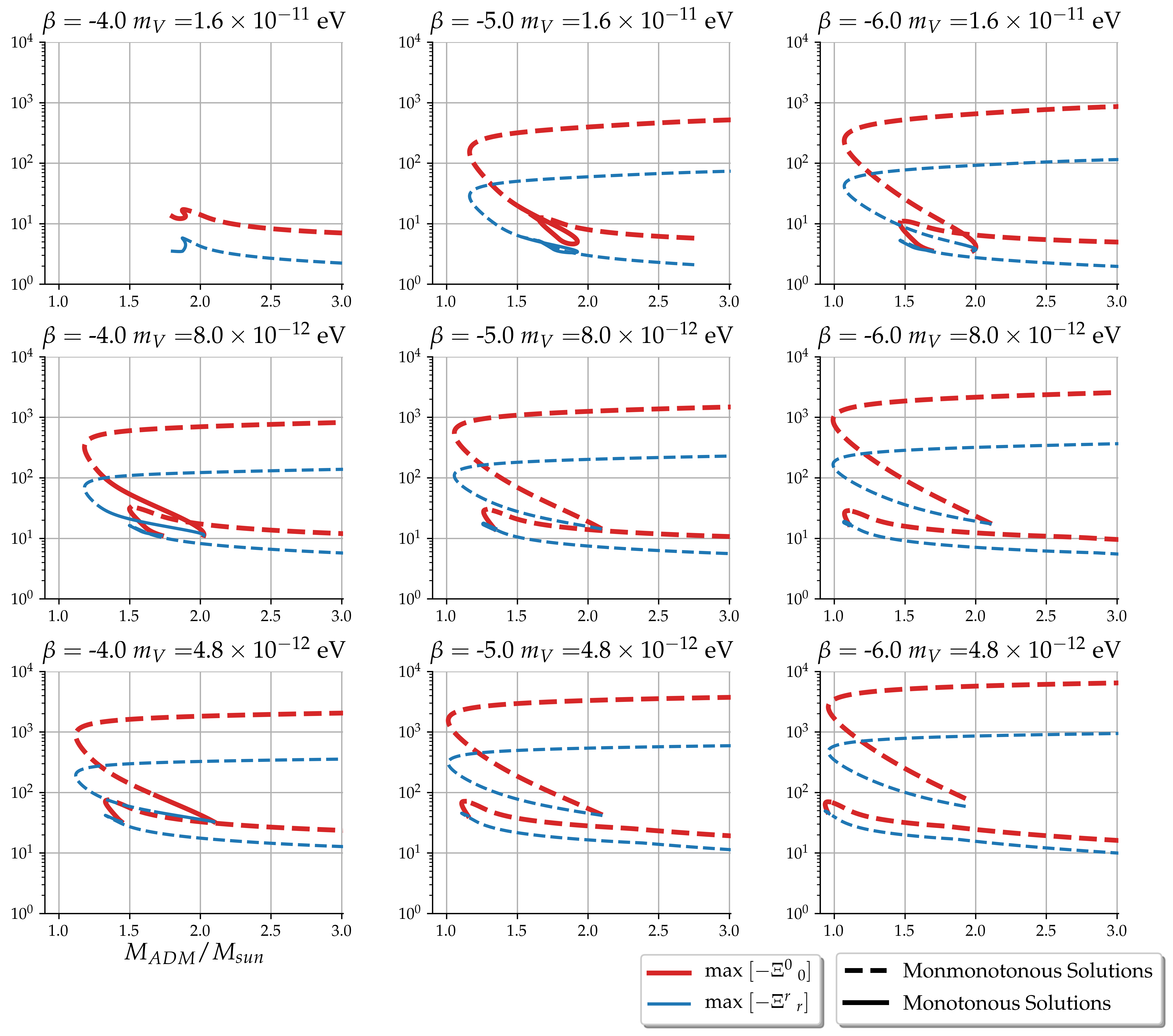}
    \caption{Maximum values of $-\Xi^0{}_0$ and $-\Xi^r{}_r$ for the neutron stars in Fig.~\ref{fig:vector_strength_negative_beta} (recall from Eq.~\eqref{eq:xiii} that $\Xi^i{}_i$ is the same for all $i$). Negative values are attained for both functions in all cases, hence all solutions have ghost and gradient instabilities.}
    \label{fig:negativebeta_Xi00}
\end{figure*}
For our purposes, we re-computed the static and spherically symmetric vectorized neutron star solutions of Ref.~\cite{Ramazanoglu:2017xbl} using the same numerical methods, and constructed other ones in a broader part of the $(\beta,\mv)$ parameter space. In addition, we computed $\Xi^\mu{}_\nu$ (Eqs.~\eqref{eq:xi00} and~\eqref{eq:xiii}) and $V_{\rm eff}$ (Eq.~\eqref{eq:Veff}) for each solution to assess stability. We checked the numerical convergence of our results, and also performed independent residual analysis. All the neutron star solutions are for the piecewise polytropic HB equation of state for nuclear matter defined in Ref.~\cite{Read:2008iy}. 

We categorize the solutions into two groups. The first group has strictly decreasing \(A_0\) and matter density \(\tilde{\rho}\), so there is no immediate sign of an instability when one looks at the solutions, hereafter ``monotonic stars". The second group have visible indications of instability, particularly \(A_0\), \(\tilde{\rho}\) or both increase at some point within the star, hereafter ``non-monotonic stars". However, note that non-monotonicity does not necessarily imply instability, nor does monotonicity imply stability, hence we examined the perturbations around all solutions. 

Summarizing the previous section, we deduce the instabilities of linearized perturbations of vectorized stars as follows:
\begin{itemize}
    \item There is a ghost instability if $\Xi^0{}_0<0$ anywhere.
    \item There is a gradient instability if $\Xi^i{}_i<0$ anywhere.
    \item There is a tachyonic instability if the eigenfunction problem in Eq.~\eqref{eq:eigenfunction} has a negative eigenvalue, $\omega^2< 0$, for $V_{\rm eff}$. As a shortcut, $u_4$ does not have a  tachyon if $V_{\rm eff}>0$ everywhere, hence such a case is an indication of stability in this mode. However, recall that this does not rule out the existence of tachyons in the $u_{1,2,3}$ modes.
\end{itemize}

As we mentioned before, having all instabilities together would mean stability if the vector field was not coupled to any other field, which is not the case in our action~\eqref{eq:action}. Moreover, issues with $\hat{z}$ cutting $0$ would remain, and the instabilities would catch up with us at higher orders in perturbation theory. Hence the above classification.

The general behavior of vectorized stars in terms of their deviation from GR can be seen in Fig.~\ref{fig:vector_strength_negative_beta}, some sample star profiles can be seen in Fig.~\ref{fig:star_profiles}.
Applying the above instability criteria to the neutron stars in Fig.~\ref{fig:vector_strength_negative_beta}, we found that all vectorized solutions have both ghost and gradient instabilities. Therefore neither type of star is stable, which is the main result of this study. The ``strength'' of the ghost and gradient instabilities of the stars as measured by the most negative values attained by $\Xi^\mu{}_\nu$ can be seen in Fig.~\ref{fig:negativebeta_Xi00}. $\Xi^\mu{}_\nu$ is significantly negative, so the existence of the ghost can not be avoided by small changes to the stellar structure. The behavior of $\Xi^\mu{}_\nu$ and $V_{\rm eff}$ as functions of radius for two sample vectorized stars can be seen in Fig.~\ref{fig:star_profiles}.

In contrast to ghosts, we find that the monotonic stars have \(V_\mathrm{eff}>0\) everywhere, hence the $u_4$ mode is not tachyonic in this case. The $u_4$ tachyon is also stabilized in the majority of the non-monotonic stars due to the same criterion. There are some non-monotonic stars where \(V_\mathrm{eff}<0\) in small regions, but we did not solve the eigenfunction problem, so these stars may or may not have tachyonic instabilities.

The naive expectation of a tachyon first growing and eventually getting quenched in vectorization seems to be partially realized for many, possibly all, solutions for the $u_4$ modes. However, note that $u_4=0$ for the vectorized star, hence, it is not the $u_4$ modes themselves that quench the tachyon, but possibly the indirect nonlinear effects of the other modes through their coupling. We reiterate that regardless of the fate of the tachyon(s), the ghost and gradient instabilities not considered in the original work~\cite{Ramazanoglu:2017xbl} are present in all solutions, which is enough to render the vectorized stars unstable. 
\begin{figure}
    \includegraphics[width=0.49\textwidth]{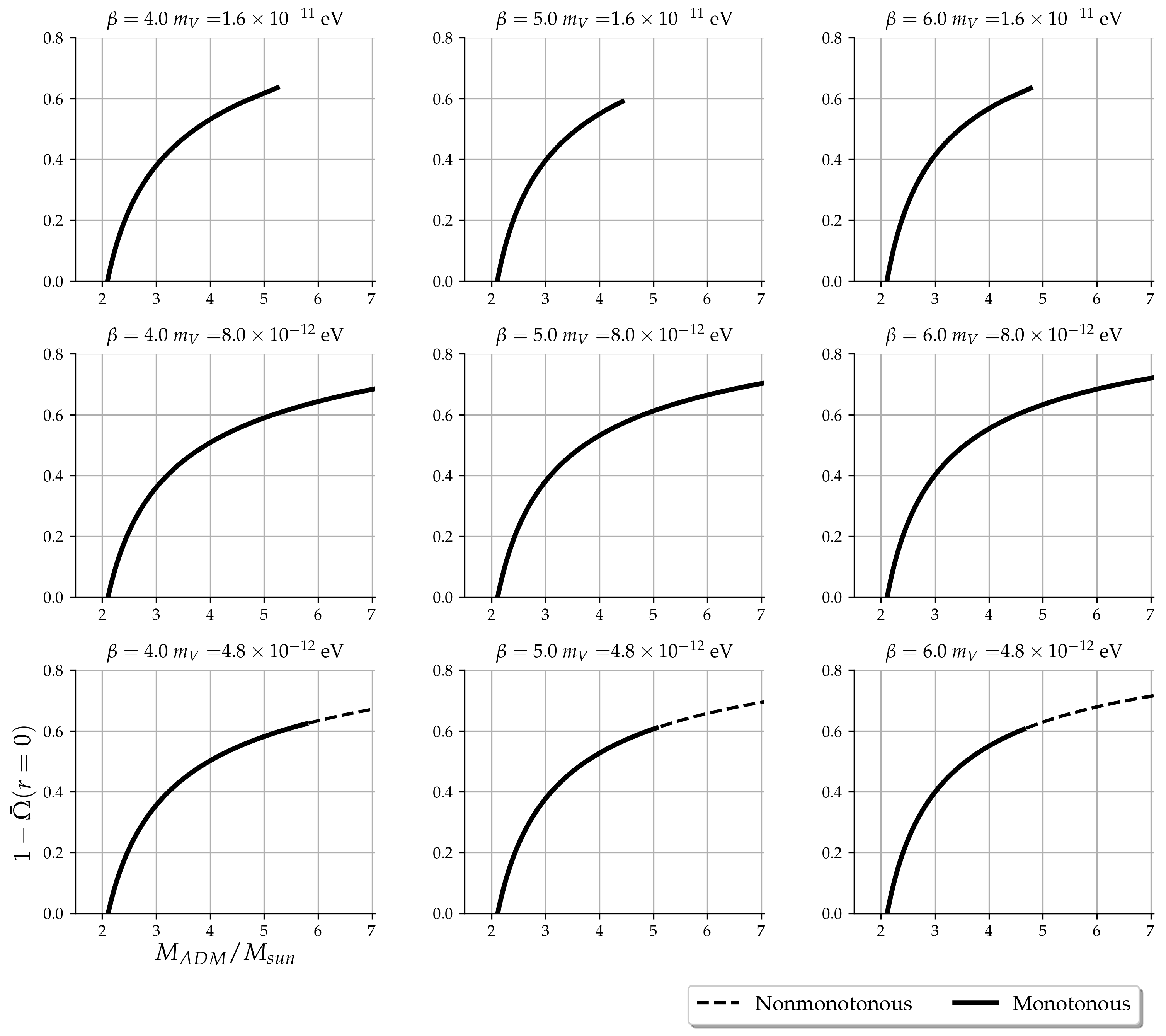}
    \caption{Deviations from GR measured by $1- \overline{\Omega}_{\rm V}(r=0)$ as a function of the ADM mass of vectorized stars for various values of $\beta>0$ and $\mv$. The behavior of the curves are relatively simpler compared to the $\beta<0$ case in Fig.~\ref{fig:vector_strength_negative_beta}, which we discuss in Sec.~\ref{sec:conclusions}. There is a second branch of solutions appearing at high $\beta$ which consists of non-monotonic solutions. All curves possibly continue to higher $M_{ADM}$ values than shown, however the numerical methods we employ are ineffective in this regime.}
    \label{fig:vector_strength_positive_beta}
\end{figure}
\begin{figure}
    \includegraphics[width=0.49\textwidth]{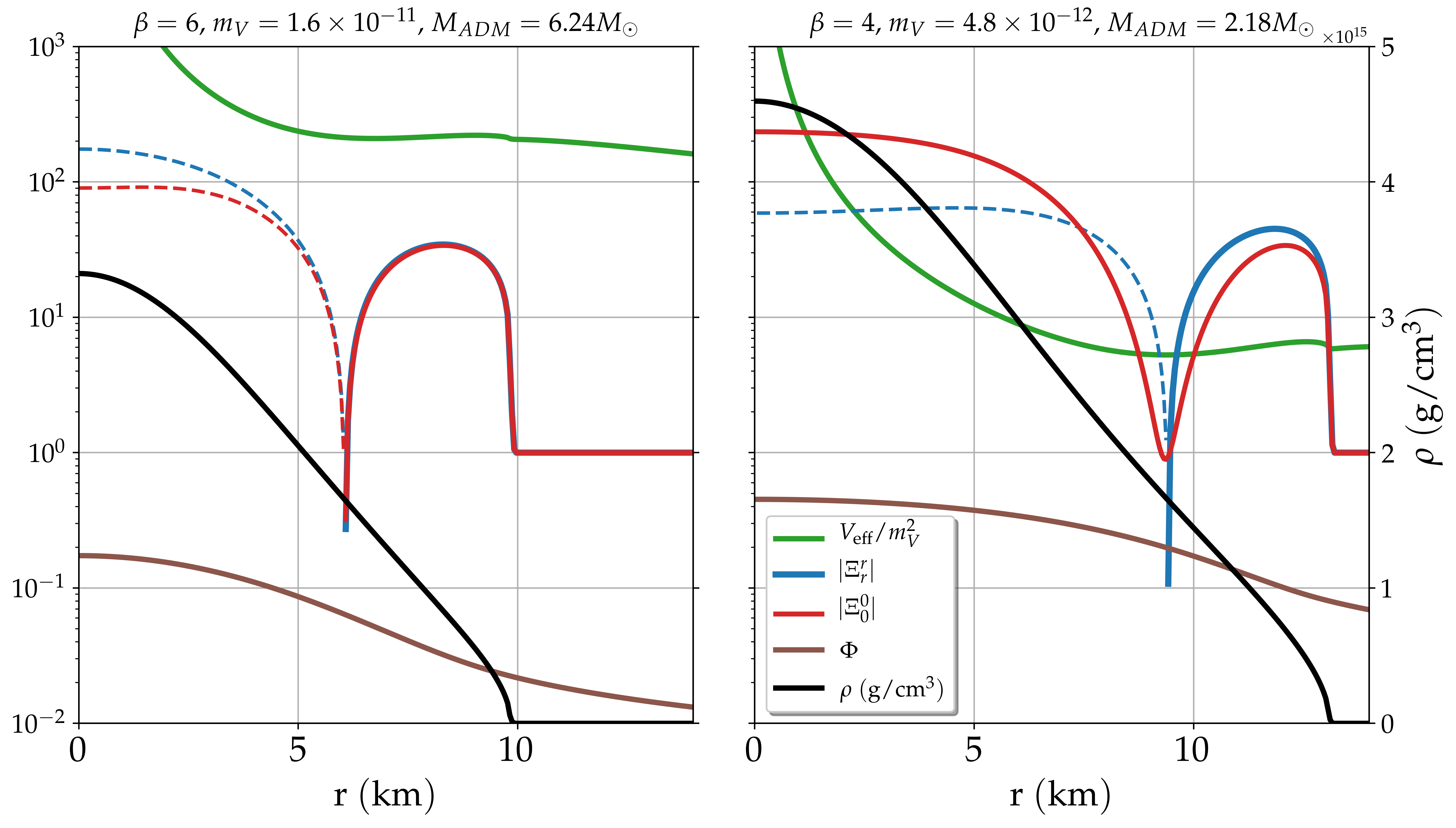}
    \caption{$\Xi^0{}_0$, $\Xi^r{}_r$, $V_{\rm eff}$, $\Phi = \sqrt{-A_0A^0}$ and the energy density $\tilde{\rho}$ of a star with (left) and without (right) ghosts for $\beta>0$. The radii where $\Xi^0{}_0$ and $\Xi^r{}_r$ cut through $0$ can be seen as the divergence in the logarithmic scale, and the lines are dashed when the functions attain negative values. Note that $\Xi^0{}_0$ and $\Xi^r{}_r$ can cut through zero twice, being negative between these radii. $V_{\rm eff}$ is not monotonically decreasing with radius in these examples, but it is positive everywhere.}
    \label{fig:star_profiles_positive_beta}
\end{figure}
\begin{figure*}
    \centering
    \includegraphics[width=0.9\textwidth]{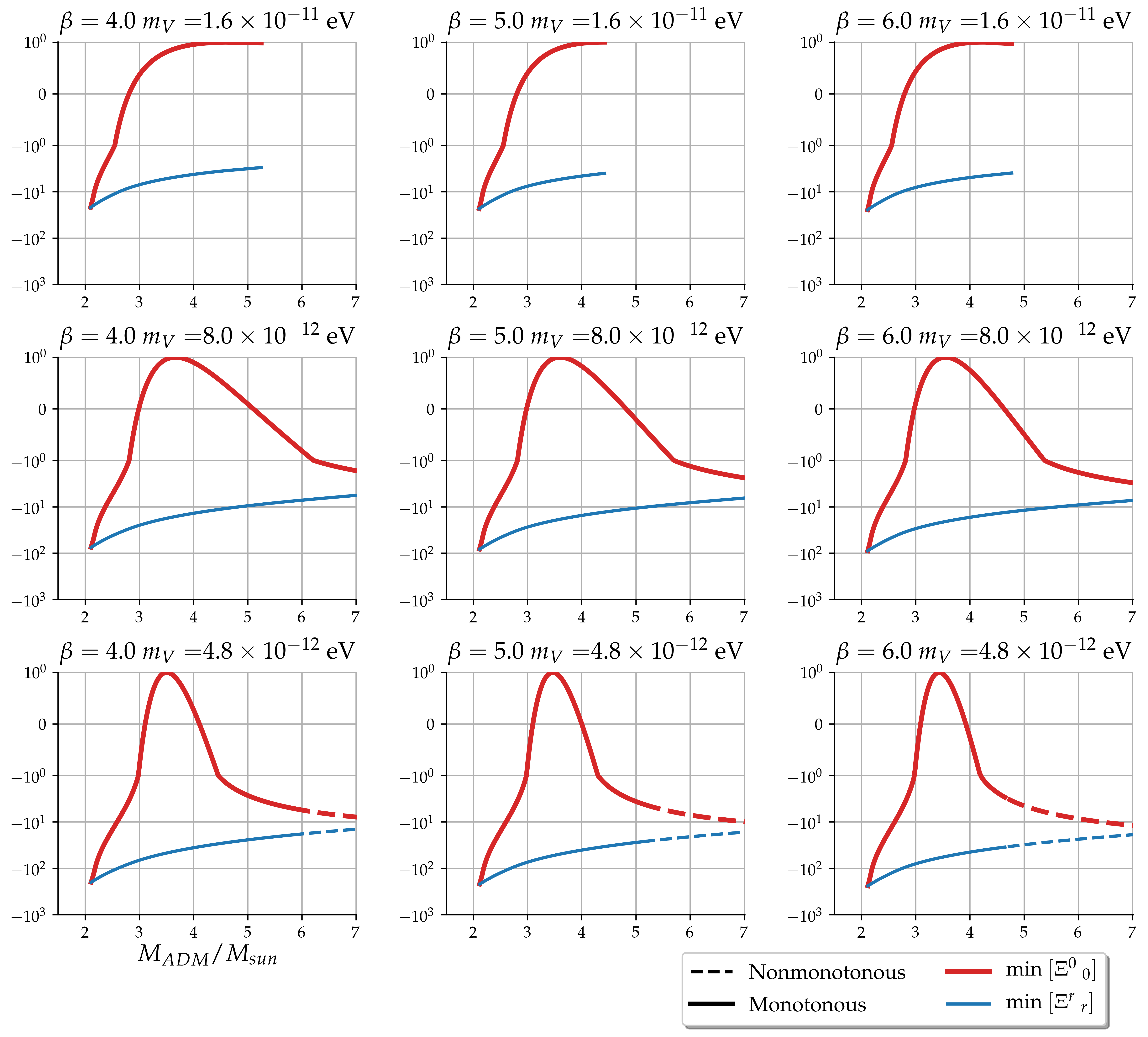}
    \caption{Maximum values of $-\Xi^0{}_0$ and $-\Xi^r{}_r$ for the neutron stars in Fig.~\ref{fig:vector_strength_positive_beta} (recall from Eq.~\eqref{eq:xiii} that $\Xi^i{}_i$ is the same for all $i$). While negative values are attained for $\Xi^r{}_r$ in all cases, $\Xi^0{}_0>0$ everywhere for some solutions. Thus, all solutions are unstable due to gradient instabilities, but some do not carry ghosts.}
    \label{fig:positivebeta_Xi00}
\end{figure*}

The solutions found in Ref.~\cite{Ramazanoglu:2017xbl} all had \(\beta<0\), however it is known, from the scalarization case, that it is possible to destabilize massive GR stars for \(\beta>0\) \cite{Mendes:2016fby}. Therefore we have investigated a part of the \(\beta>0\) parameter space for the first time as well.
The deviation from GR in this case can be seen in Fig.~\ref{fig:vector_strength_positive_beta}, monotonic and non-monotonic sample star profiles can be seen in Fig.~\ref{fig:star_profiles_positive_beta}.

The vectorized stars remained unstable in all cases for $\beta>0$ as well, which can be seen in Fig~\ref{fig:positivebeta_Xi00}, though the picture is slightly different. All solutions still have gradient instabilities hence the major problem of the $\beta<0$ solutions persists. The ghost instability remains active in the majority of the solutions, though there do exist monotonic solutions with $\Xi^0{}_0>0$ everywhere. As for the $u_4$ tachyons, there are solutions with $V_{\rm eff}>0$ and $V_{\rm eff}<0$ among both monotonic and non-monotonic stars, but we did not explore whether any of the latter cases indeed have tachyonic instabilities, since $\Xi^{i}{}_i<0$ already ensured that all solutions are unstable.

The shooting method we utilized to calculate the solutions does not work reliably for high Arnowitt-Deser-Misner (ADM) mass neutron stars, especially for higher values of $\mv$, hence the ADM masses in our figures do not necessarily reflect the most massive neutron stars in the theory. There are also solutions in parts of the parameter space where $A^0(r)$ has nodes, which are not shown in our figures. These are effectively ``higher harmonics'' of the fundamental vectorized solution where the vector field does not have nodes, and are always unstable. The solutions presented so far for very different $(\beta,\mv)$ values and ADM masses are all unstable, which strongly suggests that all static, spherically symmetric vectorized solutions are unstable, however we cannot conclusively rule out exceptions since the parameter space is vast. Further regions of the parameter space might be explored in the future using alternative numerical methods~\cite{Rosca-Mead:2020bzt}.

One final possibility we want to mention is that, however likely or not, the unstable part of the spherically symmetric regime might be dynamically avoided in vectorization, and one might have to look for the stable solutions under less stringent conditions, e.g. axisymmetry. This is possible as the non-perturbative behaviour is triggered as \(\hat{z}\) gets sufficiently small, before it crosses \(0\). Consider the concrete example of a star, far from the regime where \(\hat{z}\) crosses \(0\), collapsing to form something compact enough that, if it were to follow collapse as in GR, \(\hat{z}\) would cross \(0\). Initially this should proceed  as a GR collapse with some small perturbations, however as \(\hat{z}\) gets small, and before it crosses \(0\), perturbation theory will break down due to the $\hat{z}^{-1}$ terms. From that point on we cannot say what will happen. It is merely plausible that the non-perturbative evolution avoids the catastrophic regime we have discussed here and in Ref.~\cite{Silva:2021jya}, ending at some stable axisymmetric star. Investigation of such a scenario requires the simulation of collapsing stars, which have been performed for spontaneous scalarization~\cite{Sperhake:2017itk}.

\section{Conclusions}
\label{sec:conclusions}
Spontaneous vectorization was motivated by the fact that scalar nature of the field does not play any direct role in the DEF model. This naively presented the possibility that scalarization is not an isolated mechanism for a specific theory, but has analogs in other alternative theories of gravity. While this may be the case for more general scalar-tensor theories \cite{Andreou:2019ikc}, all indications are that the mechanism cannot be generalized to non-scalar fields. Specifically, the recent discovery of the existence of ghost and gradient instabilities around GR solutions of generic spontaneous vectorization theories showed that vectorization is radically different from scalarization~\cite{Garcia-Saenz:2021uyv, Silva:2021jya}. Even though vectorization superficially seems to be about updating the type of the field in spontaneous scalarization, the essence of the theory is about changing the type of the instability carried by the field, albeit unintentionally. 

In this work, we showed that the ghost and gradient instabilities that are present for unvectorized (GR) objects plague the static and spherically symmetric vectorized neutron star solutions as well. This is in stark contrast to the standard case of scalarized stars in the DEF model and other scalarization theories inspired by it, where the tachyonic instability is eventually quenched, leading to stability~\cite{Harada:1998ge,Salgado:1998sg,Novak:1998rk,Mendes:2016fby,Ramazanoglu:2016kul,East:2021bqk}. Our results were achieved by linearizing the vector field equations on a fixed vectorized neutron star background, and showing that the existence of instabilities is related to components of a tensor, $\Xi^\mu{}_\nu$, changing sign. 

The differences between the vectorized neutron star solutions shown in Fig.~\ref{fig:vector_strength_negative_beta}, and their counterparts in scalarization had already cast doubts on the stability of the former~\cite{Ramazanoglu:2017xbl}. We have confirmed these suspicions, and also identified the nature of the instabilities. One of the potentially tachyonic modes is suppressed in many of the vectorized stars in line with the original expectation, but the more problematic ghost and gradient instabilities are still present, something not foreseen when vectorization was introduced.

Another striking feature of vectorization is that the deviations from GR in the $\beta>0$ solutions in Fig.~\ref{fig:vector_strength_positive_beta} which are studied for the first time have a much ``simpler'' structure compared to the $\beta<0$ case in Fig.~\ref{fig:vector_strength_negative_beta}. 
We believe this is related to the large field behavior of the conformal factor $\Omega_{\rm V}$ for the vectorization of spherically symmetric stars behaving differently from that of $\Omega_\phi$ of scalarization. 
The effective mass of both the scalar (Eq.~\eqref{eq:phi_mass}) and vector (Eq.~\eqref{eq:eom_vt}) fields depend on the conformal factors $\Omega_\phi$, $\Omega_{\rm V}$ in their respective theories, since $T = \Omega_\phi^4 \tilde{T}$ in Eq.~\eqref{eq:phi_mass}. 
In the most familiar case of scalarization with $\beta<0$, the tachyonic nature is suppressed as the scalar grows since $\Omega_\phi = e^{\beta\phi^2/2}$ becomes strictly smaller. The opposite is true for scalarization with $\beta>0$, which means $\Omega_\phi$ makes the scalar even more tachyonic as it grows. This does not necessarily lead to an instability since the nonlinear changes in $\tilde{T}$ can still quench the tachyon for appropriate coupling functions, but one can encounter unstable scalarized stars as well~\cite{Mendes:2016fby}. In short, typically the $\beta<0$ case of scalarization is simpler than that of $\beta>0$~\cite{Mendes:2016fby}.

The opposite picture is true for vectorization in a spherically symmetric spacetime where $A^0$ is the only non-vanishing component, and $A_\mu A^\mu$ is negative. Thus, the simpler case where $\Omega_{\rm V} = e^{\beta A_\mu A^\mu/2}$ is suppressed with growing $|A^0|$ is that of $\beta>0$. Similarly, $\Xi^\mu{}_\nu$ also becomes less unstable, i.e. less negative, with growing $|A^0|$ for $\beta>0$ and $\tilde{T}_{\rm bg}>0$. This means we might expect the $\beta>0$ solutions of vectorization to behave more closely to the more familiar $\beta<0$ case of scalarization. This expectation seems to be partially realized in the simplicity of Fig.~\ref{fig:vector_strength_positive_beta} compared to Fig.~\ref{fig:vector_strength_negative_beta}, and in the fact that ghost instabilities are suppressed ($\Xi^{0}{}_0>0$ everywhere) for some of the vectorized stars with $\beta>0$, unlike their $\beta<0$ counterparts (see Fig.~\ref{fig:negativebeta_Xi00} vs Fig.~\ref{fig:positivebeta_Xi00}). Nevertheless, all vectorized solutions we computed are ultimately unstable for both positive and negative $\beta$.

Even though we have considered a single theory, that of action~\eqref{eq:action}, the structure of vectorized objects has been studied in other models as well~\cite{Annulli:2019fzq,Barton:2021wfj,Oliveira:2020dru}. Extended vector-Gauss-Bonnet theories present an especially interesting case where spherically symmetric vectorized black holes are known to be entropically disfavored to Schwarzschild black holes of GR, i.e. the GR solutions have higher entropy than vectorized black holes for the same ADM mass~\cite{Barton:2021wfj}. This result is likely due to the vector field and not due to changing the coupling from matter to the Gauss-Bonnet term, since scalarized black holes in scalar-Gauss-Bonnet theories are entropically favored to those of GR~\cite{Doneva:2017bvd}, and the stability of at least some of these objects has been confirmed by numerical time evolution~\cite{East:2021bqk}. Hence, we conjecture that vectorized black hole solutions of Ref.~\cite{Barton:2021wfj} are also unstable, which can be investigated by the methods we developed here.

Our approach can be adapted more broadly to any theory where the linearized vector field equation is of the form~\eqref{eq:general_form}. Generalizations of scalarization beyond vectors also exist~\cite{Ramazanoglu:2017yun,Ramazanoglu:2018hwk,Ramazanoglu:2018tig, Ramazanoglu:2019gbz, Ramazanoglu:2019jfy,Ramazanoglu:2019jrr}, and are known to suffer from similar problems~\cite{Silva:2021jya}, the only known exception being a specific form of spontaneous spinorization~\cite{Minamitsuji:2020hpl}. Developing tools for the stability analysis of solutions in these theories is a future topic of interest. 

Finally, we should reiterate that the problem of the action~\eqref{eq:action} is not merely the instability of vectorized neutron stars arising from it. Recall that the neutron star solutions with vanishing vector fields, which are solutions of GR, also suffer from ghost and gradient instabilities~\cite{Garcia-Saenz:2021uyv,Silva:2021jya}. Hence, even if the vectorized solutions were stable, the question of how the time evolution from a neutron star with no vector field to a vectorized one occurs, if it can be defined in a meaningful way at all, would remain unanswered. On the other hand, our results show that if it is possible to have an interpretation of spontaneous vectorization of action~\eqref{eq:action} as a well-posed theory, the vectorized stars that are computed so far are not the astrophysical objects one would want to study further.

\acknowledgments
AC acknowledges financial support from the European Commission and T\"{U}B\.{I}TAK under the CO-FUNDED Brain Circulation Scheme 2, Project No. 120C081. FMR is supported by a Young Scientist (BAGEP) Award of Bilim Akademisi of Turkey and by T\"UB\.ITAK Grant No 117F295. 

\bibliography{biblio}

\end{document}